\documentclass[12pt]{article}

\usepackage{amsmath,amsthm,amsopn,amstext,amscd,amsfonts,amssymb}
\usepackage{bezier,natbib}
\usepackage{graphicx}
\usepackage{etex}
\usepackage{color}
\usepackage[latin1]{inputenc}
\usepackage{latexsym,enumerate}
\usepackage{hyperref, authblk}
\usepackage{ifthen}

\def\IR{\mathbb{R}}
\def\IN{\mathbb{N}}

\def\bmx{{\mathbf x}}
\def\bmy{{\mathbf y}}
\def\bmz{{\mathbf z}}
\def\bmu{{\mathbf u}}

\def\bmw{{\mathbf w}}
\def\bmr{{\mathbf r}}

\def\bmp{{\mathbf p}}
\def\bmq{{\mathbf q}}

\def\bma{{\mathbf a}}
\def\bmn{{\mathbf n}}
\def\bmb{{\mathbf b}}
\def\bmc{{\mathbf c}}
\def\bmd{{\mathbf d}}

\def\bmA{{\mathbf A}}

\def\bmnull{{\mathbf 0}}

\newtheorem{definition}{Definition}
\newtheorem{proposition}{Proposition}
\newtheorem{lemma}{Lemma}
\newtheorem{theorem}{Theorem}

\renewcommand{\phi}{\varphi}

\DeclareMathOperator{\ext}{\,ext}

\DeclareMathOperator{\conv}{\,conv}
\def\Dwa{D_{\bmw_\alpha}}
\def\Da{\Dwa}
\def\Prob{\text{Prob}}
\def\VaR{\text{V@R}}

\newcounter{mycomments}
\newcounter{mychanges}

\newcommand{\comm}[1]
{\ifx \underconstruction \undefined
\else\addtocounter{mycomments}{1}\textbf{\color{blue}*}\marginpar{\textcolor{blue}{\tiny$^\arabic{mycomments}$ \small #1}}\fi}

\newcommand{\nnn}
{\ifx \underconstruction \undefined
\else\addtocounter{mychanges}{1}\marginpar{ \begin{flushright}
\textbf{\checkmark} \end{flushright}}\fi}

\newcommand{\checkcomments}
{\ifthenelse{\value{mycomments} < 1 && \value{mychanges} < 1}
{}
{\marginpar{\textbf{\color{blue}There \ifthenelse{\value{mycomments} = 1}{is}{are} \ifthenelse{\value{mycomments} = 0}{no}{\arabic{mycomments}} comment\ifthenelse{\value{mycomments} = 1}{}{s} \ifthenelse{\value{mychanges} = 0}{}{and \arabic{mychanges} mark\ifthenelse{\value{mychanges} = 1}{}{s}} in the text!}}}}

\bibpunct{(}{)}{;}{a}{,}{,}

\begin{document}

\title{\bf Stochastic linear programming\\ with a distortion risk constraint}
%\title{Robust linear programming\\ with distortion risk constraints}
%\title{Uncertainty sets for robust linear optimization and data trimmed regions}
%\title{Constructing distorted sets of feasible solutions for robust linear optimization}

\author[1]{\bf Karl Mosler\thanks{The corresponding author}}
\author[2]{\bf Pavel Bazovkin}
\affil[1]{\small Department of Economic and Social Statistics, University of Cologne, Meister-Ekkehart-Str. 9, D-50923 Cologne, Germany, \url{mosler@statistik.uni-koeln.de}}
\affil[2]{\small Department of Economic and Social Statistics, University of Cologne, Meister-Ekkehart-Str. 9, D-50923 Cologne, Germany, \url{bazovkin@wiso.uni-koeln.de}}

% \date{}
\date{May 11, 2012}

\maketitle

\textbf{Abstract:}
Linear optimization problems are investigated whose parameters are uncertain.
We apply coherent distortion risk measures to capture the possible violation of a restriction.
Each risk constraint induces an uncertainty set of coefficients, which is shown to be a weighted-mean trimmed region.
Given an external sample of the coefficients, an uncertainty set is a convex polytope that can be exactly calculated.
We construct an efficient geometrical algorithm to solve stochastic linear programs that have a single distortion risk constraint.
The algorithm is available as an R-package. Also the algorithm's asymptotic behavior is investigated, when the sample is i.i.d.\ from a general probability distribution. Finally, we present some computational experience.

{\it Subject classifications:}\/ Programming: stochastic. Decision analysis: risk. Statistics: nonparametric.

% {\it Area of review:}\/ Optimization.


\section{Introduction}
\label{secintro}

Uncertainty in the coefficients of a linear program is often handled by probability constraints or, more general, bounds on a risk measure.
The random restrictions are then captured by imposing risk constraints on their violation.
Consider the linear program
\begin{equation}\label{SLP}
   \bmc^\prime \bmx \longrightarrow \min  \quad s.t. \ \tilde \bmA \bmx \ge \bmb\,,
\end{equation}
and assume that $\tilde \bmA$ is a stochastic $m\times d$ matrix and $\bmb\in \IR^m$.
This is a stochastic linear optimization problem. To handle the stochastic side conditions a \textit{joint risk constraint},
\begin{equation}
   \rho^m(\tilde \bmA \bmx-\bmb) \le \bmnull\,,  \label{jointRiskConstraint}
\end{equation}
may be introduced, where $\rho^m$ is an $m$-variate risk measure.
E.g. with $\rho^m(Y)=\Prob[Y<0]-\alpha$ the restriction (\ref{jointRiskConstraint}) becomes
\begin{equation}\label{jointProbConstraint}
   \Prob[\tilde \bmA \bmx \ge \bmb] \ge 1 - \alpha\,,
\end{equation}
and a usual \textit{chance-constrained linear program} is obtained.
Alternatively, the side conditions may be subjected to \textit{separate risk constraints},
\begin{equation}
\rho^1(\tilde \bmA_j \bmx-b_j) \le 0\,, \quad j=1\dots m\,, \label{sepRiskConstraint}
\end{equation}
with $\tilde \bmA_j$ denoting the $j$-th row of $\tilde \bmA$.
In (\ref{sepRiskConstraint}) each side condition is subject to the same bound that limits the risk of violating the condition.
A linear program that minimizes $\bmc'\bmx$ subject to one of the restrictions (\ref{jointRiskConstraint}) or (\ref{sepRiskConstraint}) is called
a \textit{risk-constrained stochastic linear program}.

For stochastic linear programs (SLPs) in general and risk-constrained SLPs in particular, the reader is e.g.\ referred to \cite{KallM10}. What we call a risk measure here is mentioned there as a \textit{quality measure}, and useful representations of the corresponding constraints are given. As most of the literature, \cite{KallM10} focus on classes of SLPs with chance constraints that lead to convex programming problems, since these have obvious computational advantages; see also \citet{prek95}. Our choice of the quality measure, besides its generality, enjoys a meaningful interpretation and, as will be seen later, enables the use of convex structures in the problem. 

In the case of a single constraint ($m=1$) notate
\begin{equation}
\label{eqambiguni}
   \rho(\tilde \bma'\bmx -b) \le 0\,.
\end{equation}

A practically important example of an SLP with a single risk constraint (\ref{eqambiguni}) is the \emph{portfolio selection problem}.
Let $\tilde r_1, \dots, \tilde r_d$ be the return rates on $d$ assets and notate $\tilde \bmr=(\tilde r_1, \dots, \tilde r_d)'$.
 A convex combination of the assets' returns is sought, $\tilde \bmr'\bmx= \sum_{j=1}^d \tilde r_j x_j$, that has maximum expectation under a risk constraint and an additional deterministic constraint,
\begin{equation}\label{portfolio}
   \max_{\bmx \in {\cal C}}\ E[\tilde \bmr]'\bmx, \quad s.t.\ \rho(\tilde \bmr'\bmx)\le \rho_0,\; \bmx \in {\cal C}\,,
\end{equation}
where $\rho$ is a risk measure, $\rho_0\in \mathbb{R}$ is a given upper bound of risk (a nonnegative monetary value), and  ${\cal C}\in \IR^d$ is a deterministic set which restricts the coefficients $x_k$ in some way.
For example, if short sales are excluded, ${\cal C}$ is the positive orthant in $\IR^d$. The solution $\bmx^*$ is the optimal investment under the given model. We will see that, if a solution exists, it is regularly finite and unique.
%%despite of only one stochastic constraint we are able to get a unique finite $\bmx^*$.
In our geometric approach such a solution corresponds to the intersection of
some line and a convex body that both contain the point $E[\tilde \bmr]$.
%%, which, apparently, is always a fact.\comm{There is always a finite solution!}

Regarding the choice of $\rho$, two special cases are well known. First, let $\rho(\tilde \bmr'\bmx)= \Prob[\tilde \bmr'\bmx\le - v_0]$ and $\rho_0=\alpha$. Then the optimization problem (\ref{portfolio}) says: Maximize the mean return $E[\tilde \bmr'\bmx]$ under the restrictions $\bmx \in {\cal C}$ and
\[   \VaR_\alpha(\tilde \bmr'\bmx) \le  v_0\,.
\]
That is, the \textit{value at risk} $\VaR_\alpha$ of the portfolio return must not exceed the bound $v_0$.
Second, let
\begin{equation}\label{expected shortfall}
\rho(\tilde \bmr'\bmx)= - \frac 1\alpha \int_0^\alpha Q_{\tilde \bmr'\bmx}(t) dt\,,
\end{equation}
where $Q_Z$ signifies the quantile function of a random variable $Z$. This means that the \emph{expected shortfall} of the portfolio return
is employed in the risk restriction.

In practice, $\tilde \bma$ has to be estimated from data.
If the solution of the SLP is based on a sample of observed coefficient vectors $\bma^1,\dots,\bma^n \in \mathbb{R}^d$, that is, on an \textit{external sample}, the SLP is mentioned as an \textit{empirical risk-constrained SLP}.
In other words, we assume that $\tilde \bma$ follows an \textit{empirical distribution} that gives equal mass $\frac 1n$ to some observed points $\bma^1,\dots, \bma^n\in \IR^d$.
\cite{rockur00} investigate
an empirical stochastic program that arises in portfolio choice when the expected shortfall of a portfolio is minimized. They convert the objective into a function that is convex in the decision vector $\bmx$ and optimize it by standard methods. This approach is commonly used in more recent works of these and other authors on portfolio optimization. 
% discuss the empirical risk-constrained SLP that arises in portfolio choice with the restriction (\ref{expected shortfall}) and solve it by an algorithm.

A more complex situation is investigated by \cite{BertsimasB09}, who discuss the risk-constrained SLP with arbitrary coherent distortion risk measures, which also include expected shortfall.
These allow for a sound interpretation in terms of expected utility with distorted probabilities.
For the linear restriction a so called \textit{uncertainty set} is constructed which consists of all coefficients satisfying the risk constraint.
\cite{BertsimasB09} discuss the uncertainty set that turns the SLP into a minimax problem, called \textit{robust linear program}; however they provide no optimal solution of this program.
The uncertainty set is a convex body and, as it will be made precise below in this paper, comes out to equal a weighted-mean trimmed region.
\cite{Natar09}, on the reverse, construct similar risk measures from given polyhedral and conic uncertainty sets. \cite{Pflug2006} has proposed an iterative algorithm for optimizing a portfolio using distortion functionals, on each step adding a constraint to the problem and solving it by the simplex method.
Meanwhile, many other authors have recently contributed to the development of robust linear programs related to risk-constrained optimization problems, see, e.g. \cite{NemirovskiS06}, \cite{Ben-TalEN09}  and \cite{ChenSST10}
For a review of robust linear programs in portfolio optimization the reader is referred  to \citet{FabozziHZh10}.

In this paper we contribute to this discussion in three respects:
\begin{enumerate}
\item The uncertainty set of an SLP under a general coherent distortion risk constraint is shown to be a \emph{weighted-mean region}, which provides a useful visual and computable characterization of the set.
\item An \textit{algorithm} is constructed that solves the minimax problem over the uncertainty set, hence the SLP.
\item If the external sample is i.i.d.\ from a general probability distribution, the uncertainty set and the solution of the SLP are shown to be \textit{consistent estimators} of the uncertainty set and the SLP solution.
% \item extend the analysis and the algorithm to the SLP with multiple risk constraints ($m\ge 2$). The multi-constraint problem is solved by a geometric procedure that operates in the same dimension as the single-constraint procedure does.

\end{enumerate}

The paper is organized as follows: In Section~\ref{secdrmwmtr} constraints on distortion risk measures and their equivalence to uncertainty sets in the parameter space are discussed; further these uncertainty sets are shown to be so called weighted-mean trimmed regions that satisfy a coherency property. In Section~\ref{secsolve} a robust linear program is investigated by which the SLP with a distortion risk constraint is solved.
Section~\ref{secalguni} introduces an algorithm for this program and discusses sensitivity issues of its solution. In Section~\ref{secsamplesol} we address the SLP and its solution for generally distributed coefficients and investigate the limit behavior of our algorithm if based on an independent sample of coefficients. Section~\ref{secdiscuss} contains first computational results and concludes.

\section{Distortion risk constraints and weighted-mean regions}
\label{secdrmwmtr}

Let us consider a probability space $\langle\Omega,{\cal F}, P \rangle$ and a set $\cal R$ of random variables
%% $\tilde r:\Omega\rightarrow\IR$
(e.g. returns of portfolios).
A function $\rho: {\cal R} \rightarrow \IR$ is a law invariant \textit{risk measure} if for $Y,Z \in{\cal R}$ it holds:
 \begin{enumerate}
 \item \textit{Monotonicity}: If $Y$ is pointwise larger than $Z$ then it has less risk, $\rho(Y) \le \rho(Z)$\,.
 \item \textit{Translation invariance}: $\rho(Y+\gamma) = \rho(Y) - \gamma \ \text{for all}\;\; \gamma\in \IR$\,.
\item \textit{Law invariance}: If $Y$ and $Z$ have the same distribution, $P_Y=P_Z$, then $\rho(Y)=\rho(Z)$\,.
\end{enumerate}
$\rho$ is a \textit{coherent risk measure} if it is, in addition, positive homogeneous and subadditive,
\begin{enumerate}
\setcounter{enumi}{3}
 \item \textit{Positive homogeneity}: $\rho(\lambda Y) = \lambda \rho(Y)\quad  \text{for all}\;\;  \lambda \ge 0$\,,
 \item \textit{Subadditivity}: $\rho(Y+Z) \le \rho(Y) + \rho(Z)\quad  \text{for all}\;\;  Y,Z \in {\cal R}$\,.
\end{enumerate}
The last two restrictions imply that \textit{diversification} is encouraged - a crucial property for the risk management. Distortion risk measures are essentially the same as \textit{spectral risk measures}.
For the theory of such risk measures, see e.g.\ \citet{Foellmer04}.
A function $\rho: {\cal R} \rightarrow \IR$ is said to satisfy the \emph{Fatou property} if
${\lim \inf}_{n\to\infty} \rho(Y_n) \ge \rho(Y)$ for any bounded sequence converging pointwise to $Y$.
With the notion of coherent risk measures, we reformulate a fundamental representation result of \cite{Huber81}:

\begin{proposition}
$\rho$ is a coherent risk measure satisfying the {Fatou property} if and only if there exists a family $\mathbb{Q}$ of probability measures that are dominated by $P$ (i.e.\ $P(S)= 0 \Rightarrow Q(S)=0$ for any $S\in {\cal F}$ and $Q\in \mathbb{Q}$) such that
\[
\rho(Y) = \sup_{Q\in\mathbb{Q}}E_Q(-Y)\,.
\]
\end{proposition}

We say that the family $\mathbb{Q}$ generates $\rho$. In particular, let $(\Omega, {\cal A})=(\mathbb{R}^d,{\cal B}^d)$ and $P$ be the probability distribution of a random vector $\tilde \bma$.   Huber's Theorem implies that for any coherent risk measure $\rho$ there exists a family $\mathbb{G}$ of $P$-dominated probabilities on ${\cal B}^d$ so that
\begin{align*}
 \rho(\tilde \bma'\bmx - b)\le 0 & \quad \Leftrightarrow \quad \rho(\tilde \bma'\bmx )\le  - b\\
&  \quad \Leftrightarrow \quad  \inf_{G\in \mathbb{G}} E_G(\tilde \bma'\bmx) \ge b\\
&   \quad \Leftrightarrow \quad  E_G(\tilde \bma'\bmx) \ge b \;\; \text{for all} \;\;  G\in \mathbb{G}\,.
 \end{align*}

Let us denote by $\Delta^n$ the unit simplex in $\mathbb{R}^n$,
\[ \Delta^n= \{\bmx \in \IR^n\,|\,\sum_{k=1}^n x_k = 1, x_k \ge 0\;\; \forall k\}\,.
\]
Then, if $\tilde \bma$ has an empirical distribution on $n$ given points in $\mathbb{R}^d$, any subset ${\cal Q}$ of $\Delta^n$ corresponds to a
family of $P$-dominated probabilities, and thus defines a coherent risk measure $\rho$. As an immediate consequence of Huber's theorem an equivalent  characterization of the risk constraint is obtained (see also \cite{BertsimasB09}):
\begin{proposition}\label{empHuber}
Let $\rho: {\cal R}\to \mathbb{R}$ be a coherent risk measure and let $\tilde \bma$ have an empirical distribution on $\bma^1,\dots, \bma^n\in \IR^d$. Then there exists some ${\cal Q_\rho} \subset \Delta^n$ such that
    \begin{align*}
\rho(\tilde \bma'\bmx - b) \le 0 \quad \Leftrightarrow  \quad & \bma'\bmx \ge b \;\; \text{for all}\\
                       & \bma\in {\cal U}_\rho:= \conv\{\bma \in \mathbb{R}^d\,|\, \bma=[\bma^1,\dots,\bma^n]\bmq \,|\, \bmq \in {\cal Q}_\rho\}\,.
                                                \end{align*}
\end{proposition}
Here, $\conv(W)$ denotes the convex closure of a set $W$.
Proposition~\ref{empHuber} says that a deterministic side condition $\bma'\bmx\ge b$ holding uniformly for all $\bma$ in the \textit{uncertainty set} ${\cal U}_\rho$
is equivalent to the above risk constraint (\ref{eqambiguni}) on the stochastic side condition. This will be used below in providing an algorithmic solution of the risk-constrained SLP.

\subsection{Distortion risk measures}
\label{ssecdistrm}

A large and versatile subclass of risk measures is the class of distortion risk measures \citep{Acerbi02}.
Again, let $Q_Y$ denote the quantile function of a random variable $Y$.
\begin{definition}[Distortion risk measure]\label{distortionrisk}
Let $r$ be an increasing function $[0,1]\to [0,1]$. The risk measure $\rho$ given by
\begin{equation}\label{defdistortionrisk}
\rho(Y)= -  \int_0^1 Q_{Y}(t) dr(t)
\end{equation}
is a \textit{distortion risk measure} with \textit{weight generating function} $r$.
\end{definition}

A distortion risk measure is coherent if and only if $r$ is concave. For example, with $r(t)=0$ if $t < \alpha$ and $r(t)=1$ if $t\ge \alpha$, the \textit{value at risk} $\VaR_\alpha(Y)= -  Q_{Y}(\alpha)$ is obtained, which is a non-coherent distortion risk measure. A prominent example of a coherent distortion risk measure is the \textit{expected shortfall}, which is yielded by $r(t)=t/\alpha$ if $t < \alpha$ and $r(t)=1$ otherwise. Note that with $r(t)=t$, the risk measure becomes the expectation of $-Y$. A general distortion risk measure $\rho(Y)$ can thus be interpreted as the expectation of $-Y$ with respect to a probability distribution that has been distorted by the function $r$. In particular, a concave function $r$ distorts the probabilities of lower outcomes of $Y$ in positive direction (the lower the more) and conversely for higher outcomes (the higher the less).
In empirical applications, coherent distortion risk measures other than expected shortfall have been recently used by many authors; see, e.g., \cite{AdamHL08} for a comparison of various such measures in portfolio choice.

An equivalent characterization of a coherent distortion risk measure
is that it is coherent and {comonotonic} (\cite{Acerbi02}). $\rho$ is \textit{comonotonic} if
\[
 \rho(Y+Z) = \rho(Y) + \rho(Z) \;\; \text{for all $Y$ and $Z$  that are comonotonic},
\]
i.e., that satisfy  $\big(Y(\omega) - Y(\omega^\prime)\big)\big(Z(\omega)-Z(\omega^\prime)\big) \ge 0$ for every $\omega, \omega^{\prime}\in \Omega$.
If $Y$ has an empirical distribution on $y_1,\dots, y_n\in \mathbb{R}$, the definition (\ref{defdistortionrisk}) of a {distortion risk measure}
specializes to
\begin{equation}\label{empdistortionrisk}
 \rho(Y) = -\sum_{i=1}^{n}{q_i y_{[i]}},
\end{equation}
where $y_{[i]}$ are the values ordered from above and $q_i$ are nonnegative weights adding up to $1$. (Observe that $q_i= r(y_{[\frac{n+1-i}n]})-r(y_{[\frac{n-i}n]})$.)
Then, the distortion risk measure (\ref{empdistortionrisk}) is coherent if and only if the weights are ordered, i.e.\ $\bmq\in \Delta^n_{\le}:= \{\bmq\in \Delta^n \,|\, 0\le q_1\le\dots\le q_n\}$.

\subsection{Weighted-mean regions as uncertainty sets}\label{subsec2.2}
If $\rho$ is a coherent distortion risk measure, the uncertainty set ${\cal U}_\rho$  has a special geometric structure, which will be explored now in order to visualize the optimization problem and to provide the basis for an algorithm.
We will demonstrate that ${\cal U}_\rho$ equals a so called \textit{weighted-mean (WM) region} of the distribution of $\tilde \bma$.

 Given the probability distribution $F_Y$ of a random vector $Y$ in $\mathbb{R}^d$, weighted-mean regions form a nested family of convex compact sets,  $\{D_\alpha(F_Y)\}_{\alpha\in [0,1]}$, that are affine equivariant (that is $D_\alpha(F_{AX+b})= A\,D_\alpha(F_Y) + b$ for any regular matrix $A$ and $b\in \IR^d$). By this, the regions describe the distribution with respect to its location, dispersion and shape.
Weighted-mean regions have been introduced in \cite{DyckerhoffM10} for empirical distributions, and in \cite{DyckerhoffM10a} for general ones.

For an empirical  distribution on $\bma^1, \dots, \bma^n\in \mathbb{R}^d$, a weighted-mean region is a polytope in $\mathbb{R}^d$ and defined as
\begin{equation}
\label{eqregconv}
\Dwa(\bma^1,\dots,\bma^n)=\conv\left\{\sum_{j=1}^n w_{\alpha, j}\bma^{\pi(j)}\,\Big|\,
\text{$\pi$ permutation of $\{1,\dots,n\}$}\,\right\}\,.
\end{equation}
Here $\bmw_\alpha=[w_{\alpha,1}, \dots, w_{\alpha,n}]'$ is a vector of ordered weights, i.e.\ $\bmw_\alpha \in \Delta^n_\le$, indexed by $0\le \alpha \le 1$
that for $\alpha< \beta$ satisfies
\begin{equation}\label{majorization}
\sum_{j=1}^kw_{\alpha,j}\le \sum_{j=1}^kw_{\beta,j}\,,\quad\forall k=1,\dots,n\,.
\end{equation}
Any such family of \textit{weight vectors} $\{\bmw_\alpha\}_{0\le \alpha\le 1}$ specifies a particular notion of weighted-mean regions.
There are many types of weighted-mean regions. They contain well known trimmed regions like the
zonoid regions, the expected convex hull regions and several others. For example,
\[
w_{\alpha,j}=\left\{\begin{array}{cl}
\frac{1}{n\alpha}\,&\text{if $j>n-\lfloor n\alpha\rfloor$,}\\[1ex]
\frac{n\alpha-\lfloor n\alpha\rfloor}{n\alpha}\,&\text{if $j=n-\lfloor n\alpha\rfloor$,}\\[1ex]
0\,&\text{if $j<n-\lfloor n\alpha\rfloor$,}
\end{array}\right.
\]
$0\le \alpha\le 1$, defines the \textit{zonoid regions}.
However some popular types of trimmed regions, such as Mahalanobis or halfspace regions, are no weighted-mean regions.

 A WM region is characterized by its projections on lines. Note that each $\bmp\in S^{d-1}$, where $S^{d-1}$ is a $(d-1)$-variate unit sphere, yields a projection of the data $\bma^1,\dots, \bma^n$ on the line generated by $\bmp$ and thus induces a  permutation $\pi_\bmp$ of the data,
\[
\bmp^\prime \bma^{\pi_\bmp(1)}\le \bmp^\prime \bma^{\pi_\bmp(2)}\le\dots\le \bmp^\prime \bma^{\pi_\bmp(n)}\,.
\]
The permutation is not necessarily unique, and let $H(\bma^1,\dots, \bma^n)$ denote the set of all directions $\bmp\in S^{d-1}$ that induce a non-unique permutation $\pi_\bmp$.
\cite{DyckerhoffM10} have shown that the support function $h_\alpha$ of $\Dwa=\Dwa(\bma^1,\dots,\bma^n)$ amounts to
\begin{equation}\label{supportWMT}
    h_\alpha(\bmp)=\sum_{j=1}^nw_{\alpha,j}\bmp^\prime \bma^{\pi_\bmp(j)}\,, \quad \bmp\in S^{d-1}\,.
\end{equation}

It follows that, whenever $\pi_\bmp$ is unique, the polytope $\Dwa$ has an extremal point in direction $\bmp$,
which is given by
\begin{equation}
\label{eqext}
\sum_{j=1}^nw_{\alpha,j}\bma^{\pi_\bmp(j)} = \sum_{i=1}^n w_{\alpha,\pi_\bmp^{-1}(i)}\bma^{i}\,, \quad \bmp\in S^{d-1} \setminus H(\bma^1,\dots, \bma^n)\,.
\end{equation}

Now we are moving to the main result of this section, which will be the Theorem~\ref{thorwmtr}.
 From~(\ref{empdistortionrisk}) and (\ref{eqext}) it is seen that, with $q_i = w_{\alpha,\pi_\bmp^{-1}(i)}$ and $y_{[i]} = -\bmp'\bma^{i}$, the extreme point of the projection of $\Dwa$ on the $\bmp$-line is obtained by applying a $\bmq$-distortion risk measure to the projected data points.
Now, setting
\[{\cal Q}_\rho = \{\bmq\in \Delta^n\,|\,\bmq= (w_{\alpha,\pi_\bmp^{-1}(1)},\dots w_{\alpha,\pi_\bmp^{-1}(n)}), \bmp\in S^{d-1} \setminus H(\bma^1,\dots, \bma^n)\}
\]
and ${\cal U}_\rho= \conv \{\bma\,|\,\bma= \sum_{i=1}^n q_i \bma^{i}, \bmq\in {\cal Q}_\rho\}$ we obtain that all extreme points of $\Dwa$ are in
${\cal U}_\rho$, hence $\Dwa\subseteq {\cal U}_\rho$. On the other hand, for every $\bmq\in {\cal Q}_\rho$ it holds that
$\sum_{i=1}^n q_i \bma^{i}\in \Dwa$, which implies ${\cal U}_\rho \subseteq \Dwa$. We conclude ${\cal U}_\rho = \Dwa$.

Thus we have proven the equality between the distortion risk constraint feasible set and a properly chosen WM region, which is a
  reformulation of Theorem 4.3 from \cite{BertsimasB09}:

\begin{theorem}
\label{thorwmtr}
\begin{equation}
\label{eqrobmodel}
\{\bmx\in\IR^d\,|\, \rho(\tilde{\bma}^\prime \bmx-b)\le0\} \\
=\{ \bmx\in\IR^d\,|\, {\bma}^\prime \bmx\ge b\ \forall \bma\in \Da(\bma^1,\dots,\bma^n) \},
 \end{equation}
where $\bma^1,\dots,\bma^n$ is an external sample of the parameter vector $\tilde{\bma}$.
\end{theorem}

Recall that $\Dwa(\bma^1,\dots, \bma^n)$
is a $d$-dimensional \textit{convex polytope}, and thus the convex hull of a finite number of points (its vertices) or, equivalently,
a bounded nonempty  intersection of a finite number of closed halfspaces (that contain its facets). By this the calculation and representation of such a polytope can be done in two ways: either by its vertices or by its facets.
Recall that a nonempty
intersection of the polytope's boundary with a hyperplane is a
\textit{facet}
if it has an affine dimension $d-1$, and a \textit{ridge} if it has
an affine dimension $d-2$. It is called an \textit{edge} if it is a
line segment, and a \textit{vertex} if it is a single point.
In general, each facet of a polytope in $\IR^d$ is itself a polytope of dimension $d-1$ and has at least $d$ vertices. With WM regions the number of a facet's vertices can vary considerably; it ranges between $d$ and $d!$ \citep{BazovkinM10}. That is why in calculating WM regions a representation by facets is preferable.

The vertices of a polytope are its extreme points. From above we know that the
directions $\bmp\in S^{d-1} \setminus H(\bma^1,\dots, \bma^n)$ belong to vertices, while the
directions $\bmp\in H(\bma^1,\dots, \bma^n)$ belong to parts of the boundary that have affine dimension $\ge 1$.

In the context of risk measurement it is crucial that the WM regions  possess two properties that enable them to generate coherent risk measures: \emph{monotonicity} and \emph{subadditivity}.
\begin{proposition}{\bf (Coherency properties of WM regions)}
\begin{enumerate}
 \item Monotonicity: If $\bmz_k\le \bmy_k$  holds for all $k$ (in the componentwise ordering of $\IR^d$), then
\begin{eqnarray*}
&& \Da(\bmy_1,\dots,\bmy_n) \subset \Da(\bmz_1,\dots,\bmz_n) \oplus \IR^d_+ \,,\quad \text{and} \\
&& \Da(\bmz_1,\dots,\bmz_n) \subset \Da(\bmy_1,\dots,\bmy_n) \oplus \IR^d_- \,.
\end{eqnarray*}
 \item Subadditivity: \[ \Da(\bmy_1+\bmz_1,\dots,\bmy_n+\bmz_n) \subset  \Da(\bmy_1,\dots,\bmy_n) \oplus \Da(\bmz_1,\dots,\bmz_n)\,.
\]
\end{enumerate}
\end{proposition}
In this Proposition the symbol $\oplus$ is the \textit{Minkowski addition}, $A\oplus B = \{a+b|a\in A,b\in B\}$
  for $A$ and $B\subset\IR^d$. For a proof, see \cite{DyckerhoffM10}.

\begin{figure}[t!]
    \centering
        \includegraphics[width=0.5\textwidth]{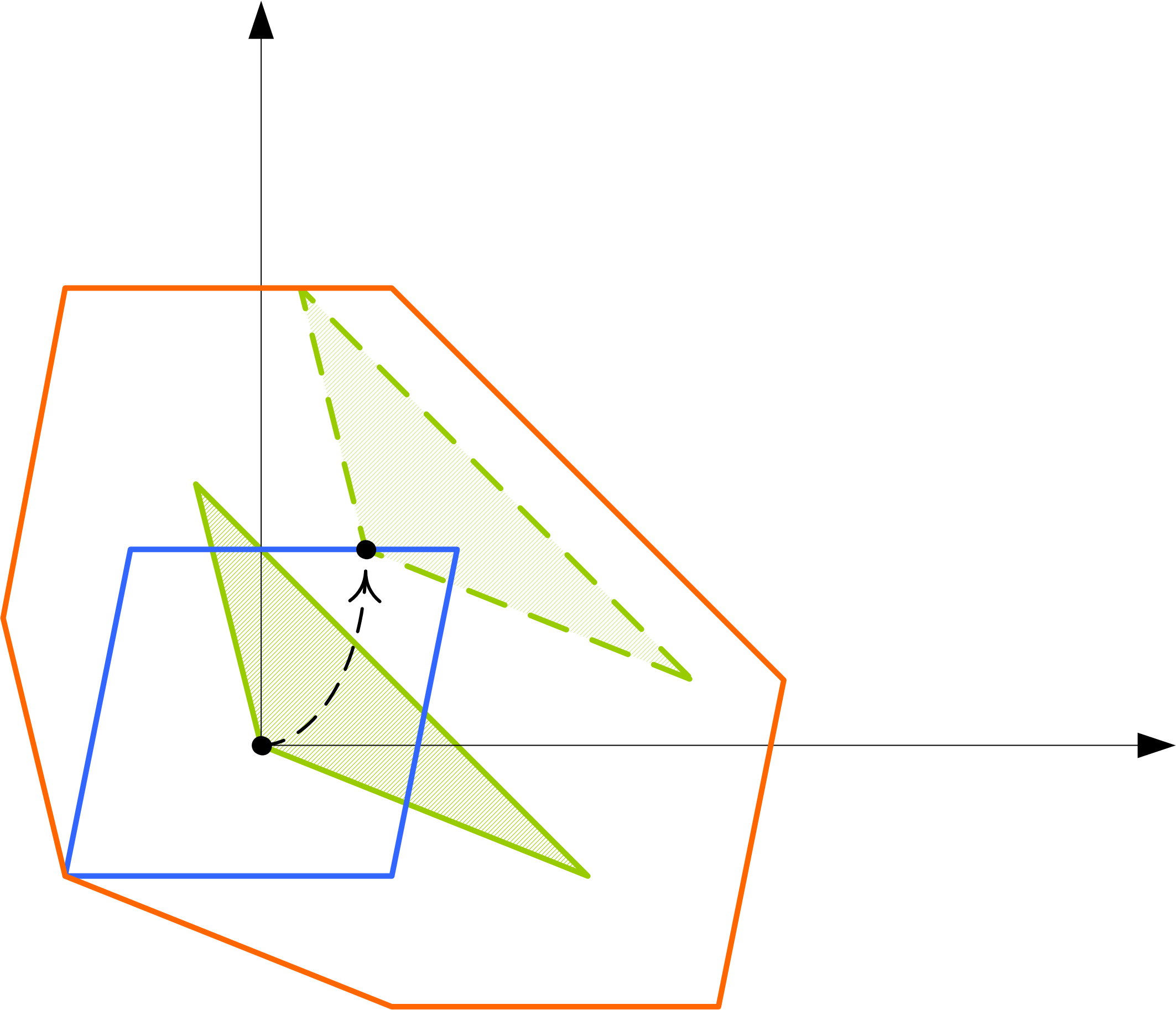}
    \caption{An illustration of the subadditivity property.}
        \label{figminkowskiex}
\end{figure}

  The subadditivity property of WM regions is an immediate extension of the subadditivity restriction usually imposed on univariate risk measures.
  In dimensions two and more it has an interpretation as a dilation of one trimmed region by the other.
   %(in sense of dilating dispersion).
   To understand this better let us consider the  simple example of Minkowski addition given in  Figure~\ref{figminkowskiex}. The figure exhibits a solid triangle with one vertex at the origin and a dotted-border quadrangle. Now move the triangle in a way that its lower left corner passes all  points of the quadrangle. At each point of the quadrangle we get a copy of the initial triangle (with a dashed border) shifted by coordinate of the point. The union of all these triangles gives us the Minkowski sum of the initial two sets, which is the big heptagon in the picture. Observe that, if the rectangle is moved around the triangle, the same sum is obtained. The subadditivity states that if, e.g., these two figures are WM regions $\Da(\bmy_1,\dots,\bmy_n)$ and $\Da(\bmz_1,\dots,\bmz_n)$ respectively, the $\Da(\bmy_1+\bmz_1,\dots,\bmy_n+\bmz_n)$ is contained by the heptagon.

\section{Solving the SLP with distortion risk constraint}
\label{secsolve}

\subsection{Calculating the uncertainty set}
\label{sseccalcuc}
\begin{figure}[t!]
    \centering
        \includegraphics[width=0.52\textwidth]{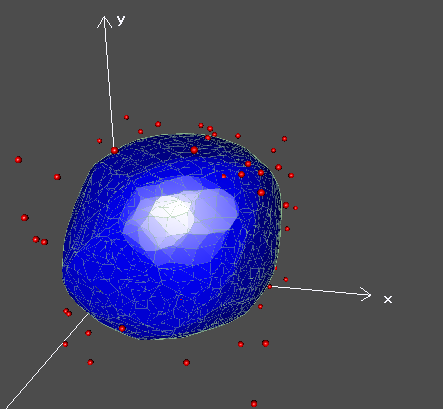},
        \includegraphics[width=0.45\textwidth]{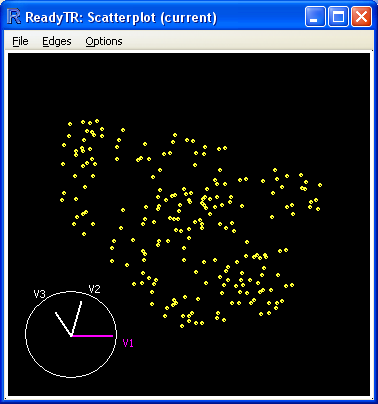}
    \caption{Visualization of WM regions by the R package \textit{WMTregions}. Left panel: Facets of a three-dimensional region in $\IR^3$. Right panel: Vertices of a four-dimensional region projected on a subspace of $\IR^3$.}
    \label{f_viszon3d}
\end{figure}

In the previous section we have shown that the uncertainty set ${\cal U}_\rho$ equals the weighted-mean (WM) region $\Dwa$ for a properly chosen weight vector $\bmw_\alpha$. \citet{BazovkinM10} provide an algorithm by which this WM region can be exactly calculated in any dimension $d$.
%%The algorithm is implemented as an R package ({\url{http://www.r-project.org/}}), \textit{WMTregions} \citep{WMTregionspack}.
%%The R package allows representing a WM region either by its vertices or by an intersection of halfspaces, giving for each of them the normal vector and the %%intercept.
The results can be visualized in dimensions two and three; for examples see Figure \ref{f_viszon3d}.

It has been also shown in \citet{BazovkinM10} that the number of vertices of a facet can be as much as $d!$ .
Therefore the representation of a WM region by its vertices appears to be less efficient than that by its facets.
In the sequel, we will use the facet representation for solving the SLP.

\subsection{The robust linear program}
\label{ssecroblp}
The robust linear program to be solved is
\begin{equation}\label{LProbust}
   \bmc^\prime \bmx \longrightarrow \min  \quad s.t. \ \bma'\bmx \ge b\;\; \text{for all } \bma\in{\cal U}\,,
\end{equation}
where the subscript $\rho$ has been dropped for convenience. The side condition is rewritten as
\begin{equation}\label{sidecond}
\bmx \in {\cal X}= \bigcap_{\bma\in {\cal U}} X_\bma \,, \quad X_\bma= \{\bmx|\bma' \bmx \ge b\}\,.
\end{equation}
Note that ${\cal X}$, as a weighted-mean region, is a convex polyhedron. So, a linear goal function is to be minimized on a convex polyhedron. Obviously, any optimal solution will lie on the surface of ${\cal X}$.
% Further on, for convenience, we will assume that $b>0$, since if $b<0$, its sign can be changed by replacing $\cal U$ through $-{\cal U}$.

\subsection{Finding the optimum on the uncertainty set}
% \subsection{Constructing a set of feasible solutions}
\label{secoptsol}
In constructing an algorithm for the robust linear program, we will explore the set ${\cal X}$ of feasible solutions and relate it to the uncertainty set ${\cal U}$ in the parameter space. It will come out that the space of solutions $\bmx$ and the space of coefficients $\bma$ are, in some sense, \textit{dual} to each other. The following two lemmas provide the connection between ${\cal X}$ and ${\cal U}$. First we demonstrate that ${\cal X}$ is the intersection of those halfspaces whose normals are extreme points of ${\cal U}$.

\begin{lemma}%%(Transition from the parameter to the solution space)
\label{ldualparamtodata}
It holds that
\[
   {\cal X}  = \bigcap_{\bma\in {\cal U}}\{\bmx\,|\, \bma' \bmx \ge b\}  = \bigcap_{\bma\in {\ext{\cal U}}}\{\bmx\,|\, \bma' \bmx \ge b\}\,.
\]
\end{lemma}

\textbf{Proof.} We show that $\bigcap_{\bma\in {\ext{\cal U}}}{X_\bma} \subset X_{\bmu}$ for all ${\bmu}\in {\cal U}$; then
$\bigcap_{\bma\in{\ext{\cal U}}}{X_\bma} \subset \bigcap_{\bma\in {\cal U}}{X_\bma}\,.$ The opposite inclusion is  obvious.
  Assume ${\bmu}\in {\cal U}$. Then, as ${\cal U}$ is convex and compact, ${\bmu}$ is a convex combination of some points $\bma^1,\dots, \bma^\ell\in {\ext{\cal U}}$, i.e.  $\bmu=\sum_{i=1}^\ell \lambda_i \bma^i$ with $\lambda_i\ge 0$ and $\sum_{i=1}^\ell \lambda_i=1$, and for any $\bmx\in \bigcap_{\bma\in{\ext{\cal U}}}{X_\bma}$ holds $\bmx\in X_{\bma^j}$ and ${\bma^j}'\bmx\ge b$ for all $j$, hence
${\bmu}'\bmx=\sum_{i=1}^\ell \lambda_i {\bma^i}'\bmx\ge b$, that is, $\bmx \in X_{\bmu}$. \hfill $\qed$

Lemma~\ref{ldualparamtodata} says
that each facet of the set ${\cal X}$ of feasible solutions corresponds to a vertex of the uncertainty set ${\cal U}$.
Hence it is sufficient to consider the extreme points of the uncertainty set.

As a generalization of Lemma~\ref{ldualparamtodata}, we may prove by recursion on $k$: Each $k$-dimensional face of the feasible set corresponds to a
$(d-k)$-dimensional face of the uncertainty set in the solution space.  This resembles the dual correspondence between convex sets and their \textit{polars} (cf. e.g. \citet{Rock1997}). However, in contrast to polars, in our case the correspondence is not reflexive.

\begin{figure}[t!]
    \centering
        \includegraphics[width=1.0\textwidth]{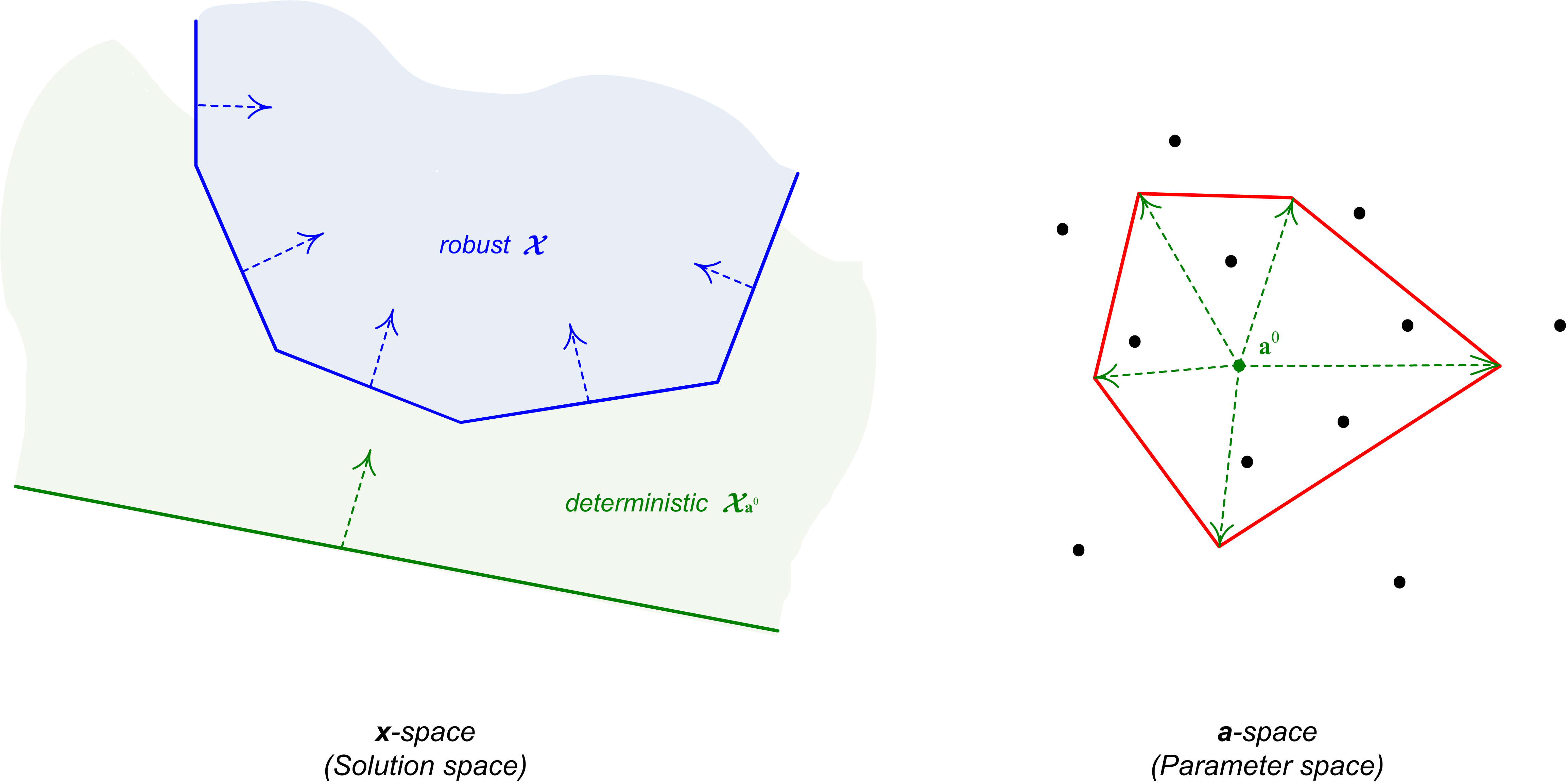}
    \caption{Deterministic and robust cases: feasible set (left panel), uncertainty set (right panel).}
        \label{fighypcurve}
\end{figure}

From Lemma~\ref{ldualparamtodata} it is immediately seen, how the robust optimization problem
contrasts with a deterministic problem, where the empirical distribution of $\tilde \bma$ concentrates at some $\bmu\in {\cal U}$.
Observe that the deterministic feasible set ${\cal X}$ is just a halfspace, ${\cal X}_\bmu=\{\bmx\,|\,{\bmu}'\bmx\ge b\}$. In the general robust case a halfspace is obtained for each $\bma\in \ext{\cal U}$, and the robust feasible set ${\cal X}$ is their intersection. The halfspaces are bounded by hyperplanes with normals equal to $\bma\in \ext{\cal U}$, and their intercepts are all the same and equal to $b$. Consequently,
the robust feasible set ${\cal X}$ is always included in the deterministic feasible set ${\cal X}_{\bmu}$,
\[ {\cal X} \subseteq  {\cal X}_{\bmu} \quad \text{for any}\;\; \bmu \in {\cal U}\,.
\]
Moreover, the two feasible sets cannot be equal unless each element of ${\cal U}$ is a scalar multiple of $\bmu$ with a factor greater than one, ${\cal U} \subseteq \{\bma\,|\,\bma = \lambda \bmu, \lambda>1\}$.
Consequently, the minimum value of the robust stochastic LP cannot be smaller than the value of an LP with any deterministic parameter $\bmu$ chosen from the uncertainty set. Figure~\ref{fighypcurve} (left panel) illustrates how a deterministic feasible set in dimension two compares to a general robust one: The line that bounds the halfspace ${\cal X}_{\bmu}$ `folds' into a piecewise linear curve delimiting ${\cal X}$.

Let
\[{U_\bmx}  = \{\bma\in \IR^d| \bma'\bmx \ge b\}\,, \quad \bmx \in \IR^d\,.
\]

\begin{lemma}%%(Transition from the solution to the parameter space)
   \label{ldualdatatoparam}
It holds that
\[
   {\cal U} \subset \bigcap_{\bmx\in {\cal X}}{U_\bmx}  \subset \bigcap_{\bmx\in \ext{\cal X}}{U_\bmx}\,.
\]
Moreover, each vertex $\bmx\in \ext{\cal X}$ corresponds to a facet of ${\cal U}$.
\end{lemma}

\textbf{Proof.}
 By Lemma~\ref{ldualparamtodata} we have $\bmx\in{\cal X} \quad \Leftrightarrow \quad \bma'\bmx\ge b$ for all $\bma\in {\cal U}$. Now let $\bma\in {\cal U}$; then for any $\bmx\in{\cal X}$ it holds that $\bma'\bmx\ge b$, hence $\bma\in U_\bmx$. Conclude ${\cal U} \subset \bigcap_{\bmx\in {\cal X}}{U_\bmx}$.
 Further, it is clear that an extreme point $\bmx\in \ext{\cal X}$ yields a facet of ${\cal U}$.
\qed

\textbf{Remark.} While ${\cal U}$ is always compact, ${\cal X}$ is in general not.
Therefore neither inclusion holds with equality.

The ordinary simplex algorithm, operating on the vertices of ${\cal X}$, constructs a chain of adjacent facets in the space of parameters. The chain ends at the solution of the optimization task. Notice that this chain corresponds to a chain of facets of the uncertainty set. So, in principle we could try to calculate this chain of facets in the parameter set. However, in our algorithm, another way is pursued to find the optimal solution.

To manage this task let us consider the goal function $\bmc'\bmx$. In the parameter space $\bmc$ corresponds to a point or a direction. In the solution space it corresponds to all hyperplanes that have $\bmc$ as their normal.
To produce all these hyperplanes in the parameter space, $\bmc$ has to be multiplied with some scaling factor. Hence the hyperplanes are obtained by passing through a straight ray $\phi$ starting at the origin and containing $\bmc$.

\begin{figure}[t!]
    \centering
        \includegraphics[width=1.0\textwidth]{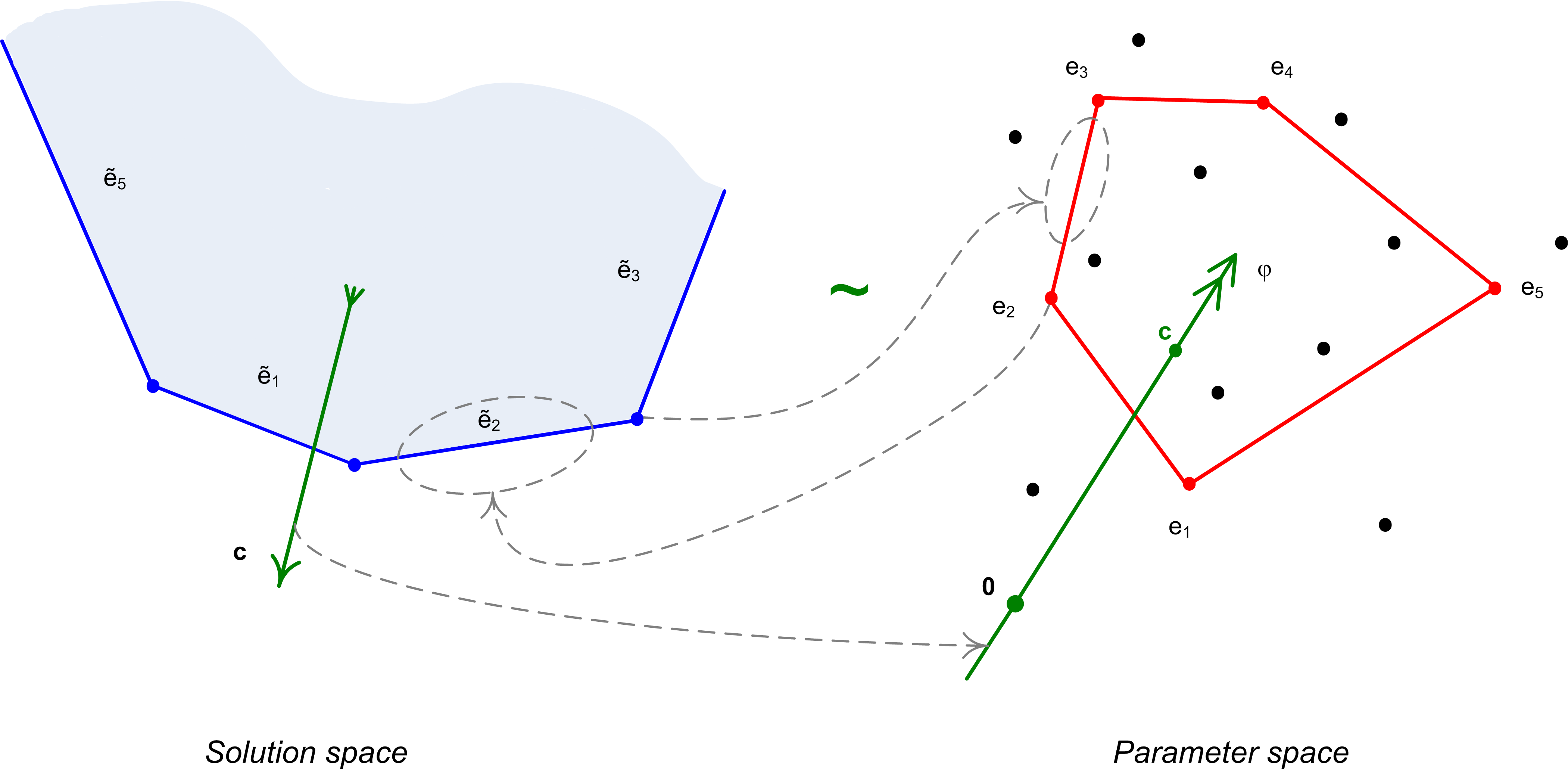}
    \caption{Duality between spaces.}
        \label{figduality}
\end{figure}

Next we search the intersection of  ${\cal U}$ with the ray $\phi$. Note that finding the intersection of a line and a polyhedron in $\IR^3$ is an important problem in computer graphics (cf. \citet{Kay1986}). The same principle is employed for a general dimension $d$.
The uncertainty set ${\cal U}$ is the finite intersection of halfspaces ${\cal H}_j$, $j=1\dots J$, each being defined by a hyperplane $H_j$ with normal $\bmn_j$ pointing into ${\cal H}_j$ and an intercept $d_j$.

Consider some point $\bmu$ on the ray $\phi$ that is not in ${\cal U}$. Compute $\frac{d_j}{\bmu'\bmn_j}$ for all halfspaces ${\cal H}_j$ that do \emph{not} include $\bmu$, i.e.\ where $(\bmu'\bmn_j - d_j)< 0$ holds. (In other words, $H_j$ is \textit{visible} from $\bmu$.) Find $j_*$ at which this value is the largest.
    Recall that moving a point $\bmu$ along $\phi$ is equivalent to multiplying $\bmu$ by some constant. The furthest move is given by the biggest constant. The \emph{optimal solution} $\bmx^*$ of the robust SLP has to satisfy
    $\bma'\bmx^*\ge b$, which is equivalent to
    \[ \bma'\left(\frac{d_{j_*}}b \bmx^*\right) \ge d_{j_*}\,.
    \]
Hence, to obtain $\bmx^*$, the normal $\bmn_{j_*}$ has to be scaled with the constant $\frac{b}{d_j}$,
\begin{equation}\label{optimalsolution}
\bmx^* = \frac{b}{d_{j_*}}  \bmn_{j_*}\,.
\end{equation}

Besides the regular situation described above, two special cases can arise:
\begin{enumerate}
   \item There is no facet visible from the origin. This means that no solution is obtained.
\item $\phi$ does not intersect $\cal U$. Then the whole procedure is repeated with the opposite ray $- \phi$. If this still gives no intersection, an infinite solution exists.
\end{enumerate}

Finally, we like to point out that
not the whole polytope ${\cal U}$ needs to be calculated but only a part of it which intersects the
ray~$\phi$. In searching for the optimum not all $F$ facets need to be checked, but only a subset of the surface where the intersection will happen. Such a filtration makes the procedure more efficient. The search for a proper subset can be driven by \textit{geometrical} considerations.
Let $\bmx^*$ be an \emph{optimal solution} of the robust SLP.   A subset ${\cal U}_\text{eff}$ of ${\cal U}$ will be mentioned as an \textit{efficient parameter set} if
\begin{itemize}
  \item $\bmx^* \in  \bigcap_{\bma\in {\cal U}_\text{eff}}\{\bmx| \bma' \bmx \ge b\ \} \subset {\cal X} \quad \text{and}$
  \item $\bma, \bmd\in {\cal U}_{\text{eff}}, \ \bma'\bmx\ge b, \ \bmd'\bmx\ge b \quad \text{implies} \quad \bma=\bmd\,.$
\end{itemize}

%%\footnote{Some $\bma_*$ is \textit{dominated} by a parameter set $\cal A$ if $\forall{\bma\in{\cal A}}$  holds $\bma^\prime\bmx\ge %%b\Rightarrow\bma_*^\prime\bmx\ge b$.}

That is to say, ${\cal U}_\text{eff}$ is the minimal subset of $\cal U$ containing all facets that can be optimal for some $\bmc$.
%%irrespective of a given $\phi$. Obviously, ${\cal U}_\text{eff}\subset\partial\cal U$ holds. Moreover:

\begin{proposition}
\label{proplowerbound}
${\cal U}_\text{eff}$ is the union of all facets of $\cal U$ for which $d_j\ge 0$ holds.
\end{proposition}
In other words, an efficient parameter set ${\cal U}_\text{eff}$ consists of that part of the surface of $\cal U$ that is visible from the origin $\mathbf 0$. The proof is obvious.

%%\textbf{Proof:} It can be easily seen, that only this part of $\cal U$ can be intersected by a line containing ${\mathbf 0}$, in our case, $\phi$. %%On the other hand, each point of ${\cal U}_\text{eff}$ can become an intersection of $\phi$ and ${\cal U}$ depending on $\phi$. The latter proves %%the minimality. \qed

\begin{figure}[t!]
    \centering
        \includegraphics[width=1.0\textwidth]{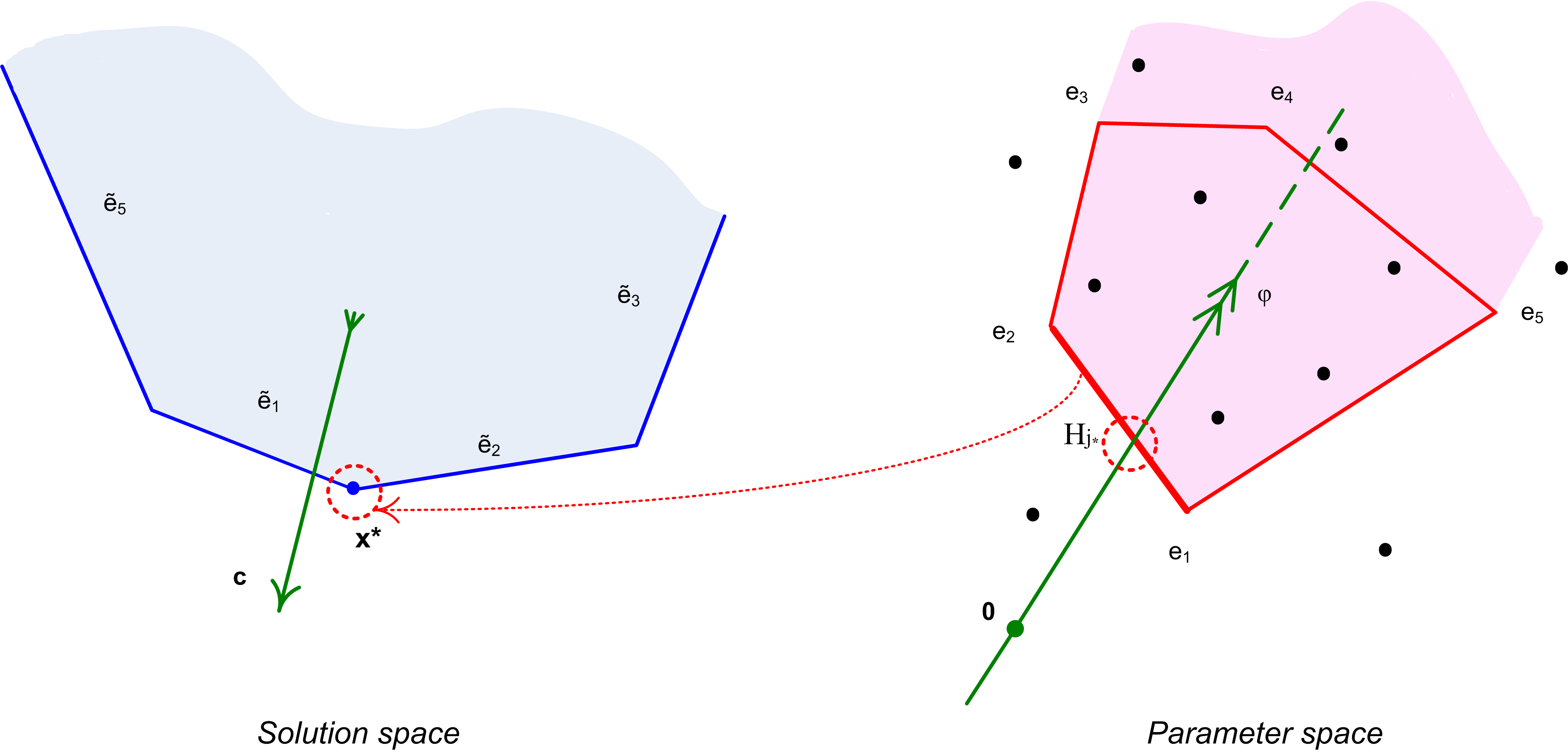}
    \caption{Finding the optimal solution on the uncertainty set.}
        \label{figalguni}
\end{figure}

To visualize the efficient parameter set we will use the \textit{augmented uncertainty set}, which is defined as
\[ \{\bma| \bma = \lambda \bma^*, \lambda>1,\bma^*\in{\cal U}_\text{eff}\}\,.
\]
It includes all parameters that are dominated by ${\cal U}_\text{eff}$; see the shaded area in the right panel of Figure~\ref{figalguni}.

%%\comm{Case $b<0$}
So far we have assumed that $b>0$. It is easy to show, that with $b<0$ we have to construct the intersection of $\phi$ with the part of the surface of $\cal U$ that is \textit{invisible} from the origin $\mathbf 0$, which is $\tilde{\cal U}_\text{eff}$ in this case. In the sense of Proposition~\ref{proplowerbound}, $\tilde{\cal U}_\text{eff}$ contains all facets of $\cal U$ with $d_j\le 0$. Obviously, $\tilde{\cal U}_\text{eff}$ is always non-empty in this case, which, in turn, means that the existence of a solution is guaranteed. However, the solution can be infinite if $\phi$ does not intersect $\tilde{\cal U}_\text{eff}$.

The situation of $b<0$ is common in the maximizing SLPs. Really, if we have the model
\begin{equation}\label{LProbust-max}
   \bmc^\prime \bmx \longrightarrow \max  \quad s.t. \ \bma'\bmx \le b\;\; \text{for all } \bma\in{\cal U}\,,
\end{equation}
it is possible to rewrite it as follows:
\begin{equation}\label{LProbust-max-trans}
   (-\bmc)^\prime \bmx \longrightarrow \min  \quad s.t. \ (-\bma)'\bmx \ge -b\;\; \text{for all } \bma\in{\cal U}\,.
\end{equation}

Clearly,~\eqref{LProbust-max-trans} is equivalent to~\eqref{LProbust} except of the negativity of the coefficient~$b$.

\section{The algorithm}
\label{secalguni}

In this part an accurate procedure of obtaining the optimal solution is given.
% \comm{Must be more precise and detailed!}

\textit{Input:}
\begin{itemize}
\item a vector $\bmc\in \IR^d$ of coefficients of the goal function,
\item an external sample $\{\bma^1,\dots,\bma^n\}\subset \IR^d$ of coefficient vectors of the restriction,
 \item a right-hand side $b\in \IR$ of the restriction,
\item a distortion {risk measure} $\rho$ (defined either by name or by a weight vector).
\end{itemize}

\textit{Output:}
\begin{itemize}
\item the {uncertainty set} ${\cal U}$ of parameters given by
\begin{itemize}
  \item facets (i.e. normals and intercepts),
  \item vertices,
\end{itemize}
\item the {optimal solution} $\bmx^*$ of the robust LP and its value $\bmc'\bmx^*$.
\end{itemize}

\textit{Steps:}
\begin{enumerate}
\renewcommand{\theenumi}{\Alph{enumi}.}   \renewcommand{\labelenumi}{\theenumi}

\renewcommand{\theenumii}{\alph{enumii}.}   \renewcommand{\labelenumii}{\theenumii}

\renewcommand{\theenumiii}{\Roman{enumiii}.}   \renewcommand{\labelenumiii}{\theenumiii}

\renewcommand{\theenumiv}{\roman{enumiv}.}   \renewcommand{\labelenumiv}{\theenumiv}

\item Calculate the subset ${\cal U}_\text{eff}\subset {\cal U}$ consisting of facets $\{(\bmn_j,d_j)\}_{j\in J}$.
\item Create  a line $\phi$ passing through the origin $\mathbf 0$ and $\bmc$.
\item Search for a facet $H_{j_*}$ of ${\cal U}_\text{eff}$ that is intersected by $\phi$:
%%%Its rescaled normal is the optimal solution $\bmx^*$ of the robust LP.
\begin{enumerate}
\item Select a subset ${\cal U}_{sel}\subseteq{\cal U}_\text{eff}$ of facets: This may be either ${\cal U}_\text{eff}$ itself or its part where the intersection is expected; ${\cal U}_{sel}=\{ (\bmn_j,d_j)\,|\,j\in J_{sel}\}$. For example, we can search the best solution on a pre-given subset of parameters. The other possible filtration is iterative transition to a facet with better criterion value.
\item
\label{stfindbest} Take a point $\bmu=\lambda \bmc, \lambda \ge 0$, outside the augmented uncertainty set. Find the $j_* = \arg\underset{j}{\max}\{\lambda_j=\frac{d_j}{\bmu'\bmn_j}|\lambda_j>0\}_{j\in J_{sel}\subseteq J}$. For the case $b<0$ just replace $\arg{\max}$ with $\arg{\min}$.
\begin{enumerate}
   \item If $\phi$ does not intersect ${\cal U}_\text{eff}$, then the solution is \textit{infinite}. If $b>0$, then repeat \ref{stfindbest} for the opposite ray $-\phi$; else stop.
\item If in the case $b>0$ the inner part of $\cal U$ contains the origin, then \textit{no solution} exists; stop.
\end{enumerate}

% \item Take a point $\bmu=\lambda \bmc, \lambda \ge 0$, outside the augmented uncertainty set. Find the $j_* = \arg\underset{j}{\max}\{\frac{d_j}{\bmu'\bmn_j}\}_{j\in J_{sel}\subseteq J}$.
\item $\bmx^* = \frac{b}{d_{j_*}}  \bmn_{j_*}$ is the optimal {solution} of the robust LP.
\end{enumerate}

\end{enumerate}

 In fact, the line $\phi$ consists of points that correspond to hyperplanes whose normal is the vector $\bmc$ in the dual space. One part of $\phi$ is dominated by points from ${\cal U}_\text{eff}$, while the other is not (which results from Proposition~\ref{proplowerbound}). The crossing point $\bma^*$ defines the hyperplane that touches the feasible set at the optimum as its dual.

Moreover, a typical nonnegativity side constraint $\bmx \ge {\bf 0}$ can be easily accounted for in the algorithm. In considering this, the search for facets has just to be restricted to those having nonnegative normals.

To solve the portfolio selection problem~\eqref{portfolio} with the algorithm, we treat the realizations of the vector $-\tilde\bmr$ of losses rates as $\{\bma^1,\dots,\bma^n\}$, and minimize $\bmc' \bmx$ with $\bmc = \frac{1}{n}\sum_{i=1}^{n}{\bma^i}$. This corresponds to transforming the maximizing SLP by \eqref{LProbust-max-trans} and running the above outlined procedure.
%%\comm{max $\to$ min}
Note that both $\phi$ and $\cal U$ contain the point $\frac{1}{n}\sum_{i=1}^{n}{\bma^i}$, that is, they always intersect, which, in turn, guarantees the existence of a finite solution.
To meet a unit budget constraint, the solution $\bmx^*$ is finally scaled down by $\sum_{j=1}^d x^*_j=1$. Recall that the risk measure is, by definition, scale equivariant.

%%It should also be noted that multiplying $\rho_0$ by a constant in~\eqref{portfolio} leads to the same scaling of the solution $\bmx^*$, that is, the relative %%weights of the assets investments do not change. In fact, due to the properties of the distortion risk measure, $\rho(\tilde \bmr'\bmx)\le \tau\cdot\rho_0$ is %%equivalent to $\rho(\tilde \bmr'\cdot(\frac{1}{\tau}\bmx))\le \rho_0$ for every positive~$\tau$.
% Comparing to \cite{Pflug2006}

\subsection{Sensitivity and complexity issues}
\label{ssecsens}

Next we like to discuss how the robust SLP and its optimal solution behave when the data $\{\bma^1,\dots,\bma^n\}$ on the coefficients are slightly changed. From (\ref{supportWMT}) it is immediately seen that the support function $h_{\cal U}$ of the uncertainty set is continuous in the data
$\bma^j$ as well as in the weight vector $\bmw_\alpha$. (Note that the support function $h_{\cal U}$ is even \textit{uniformly continuous} in $\bma^1,\dots,\bma^n$ and $\bmw_\alpha$, which is tantamount saying that the  uncertainty set ${\cal U}$ is \textit{Hausdorff continuous} in the data and the risk weights.)
Consequently, a slight perturbation of the data will only slightly change the value of the support function of ${\cal U}$, which is a practically useful result regarding the sensitivity of the uncertainty set with respect to the data. The same is true for a small change in the weights of the risk measure.

We conclude that the point $\bma^{j_*}$ where the line through the origin and $\bmc$ cuts ${\cal U}$ depends continuously on the data and
the weights. However this is not true for the optimal solution $\bmx^*$, which may `jump' when the cutting point moves from one facet of ${\cal U}$ to a neighboring one.

The theoretical complexity in time of finding the solution is compounded from the complexity of one transition to the next facet and by the whole number of such transitions until the sought-for facet is achieved. \cite{BazovkinM10} have shown that the transition has a complexity of ${\cal O}(d^2n)$. In turn, in the same paper the number of facets $N(n,d)$ of an WM region is shown to lie between ${\cal O}(n^d)$ and ${\cal O}(n^{2d})$ depending on the type of the WM region. Thus, it is easily seen, that an average number of facets in a facets chain of a fixed length is defined by the density of facets on the region's surface, $\sqrt[d]{N(n,d)}$, and is estimated by a function between ${\cal O}(n)$ and ${\cal O}(n^2)$. The overall complexity is then ${\cal O}(d^2n^2)$ up to ${\cal O}(d^2n^3)$. Notice, that the lower complexity is achieved for zonoid regions, namely when the expected shortfall is used for the risk measure.

\subsection{Ordered sensitivity analysis}
\label{ssecorder}

\begin{figure}[t!]
    \centering
        \includegraphics[width=1.0\textwidth]{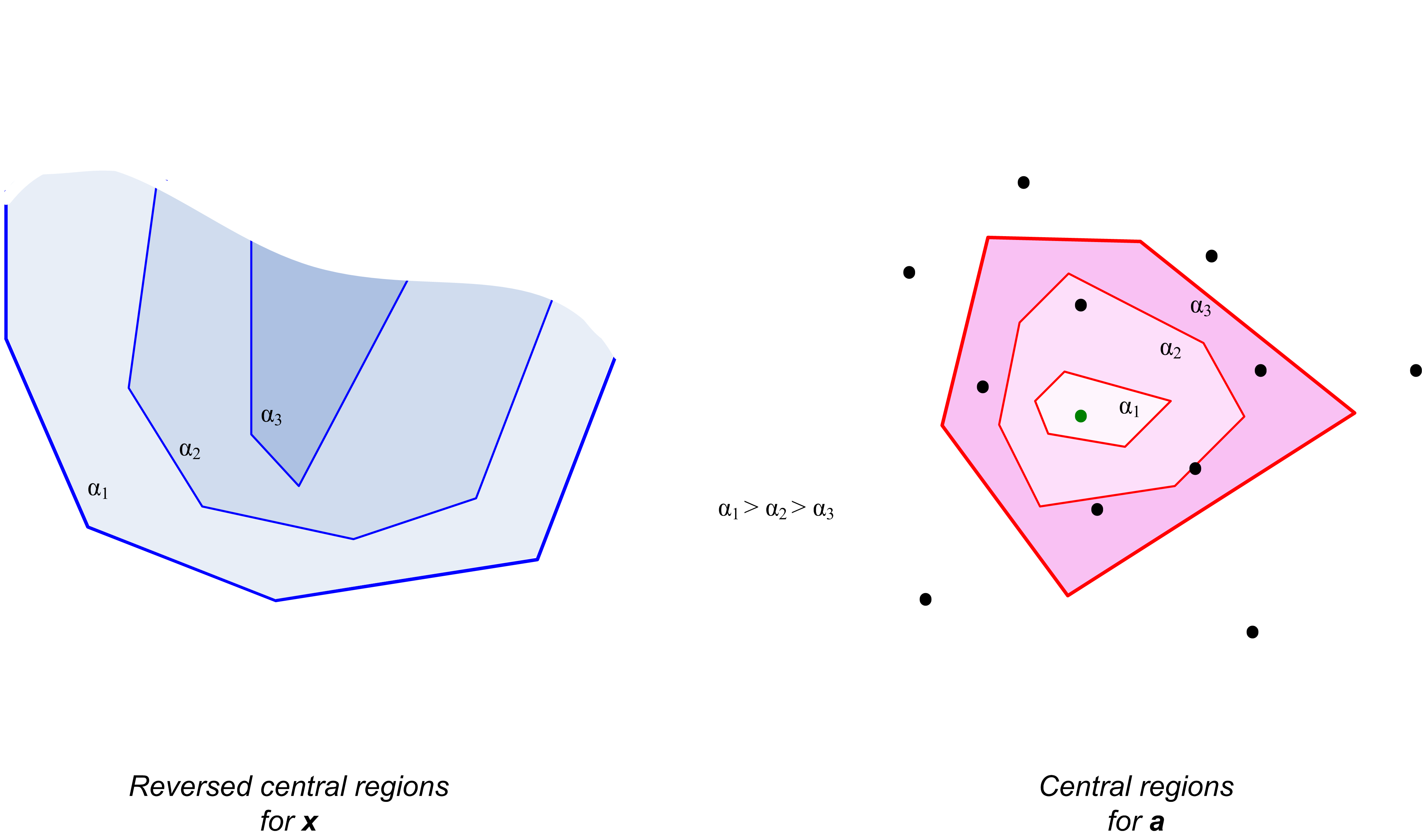}
    \caption{Example of the `reversed' central regions in the dimension 2.}
        \label{figoutofcentral}
\end{figure}

Also alternative uncertainty sets may be compared that are ordered by inclusion. From Lemma~\ref{ldualparamtodata} it is clear that the respective sets of feasible solutions are then ordered in the reverse direction; see e.g.\ Figure~\ref{figoutofcentral}. In particular we may consider the robust LP for two alternative distortion risk measures which are based on weight vectors $\bmw_\alpha$ and $\bmw_\beta$, respectively, that satisfy the monotonicity restriction (\ref{majorization}). Then the resulting uncertainty sets are nested, ${\cal U}_\beta \subset {\cal U}_\alpha$ and so are, reversely, the feasible sets, ${\cal X}_\beta \supset {\cal X}_\alpha$.
This is a useful approach for visualizing the {sensitivity} of the robust LP against changes in risk evaluation.

\section{Robust SLP for generally distributed coefficients}
\label{secsamplesol}

So far an SLP~\eqref{SLP} has been considered where the coefficient vector $\tilde \bma$ follows an empirical distribution. It has been solved on the basis of an external sample $\{\bma^1,\dots,\bma^n\}$.
In this section the SLP will be addressed with a general probability distribution $P$ of $\tilde \bma$. We formulate the robust SLP in the general case and demonstrate that the solution of this SLP can be consistently estimated by random sampling from $P$.

Consider a distortion risk measure $\rho$ (\ref{defdistortionrisk})
that measures the risk of a general random variable $Y$ and has weight generating function $r$,
$\rho(Y)= -  \int_0^1 Q_{Y}(t) dr(t)$.
Similarly as in Section \ref{subsec2.2} a convex compact $\cal U$ in $\IR^d$ is constructed through its support
function $h_{\cal U}$,
\[ h_{\cal U}(\bmp) = \int_0^1 Q_{\bmp'\tilde \bma}(t) dr(t)\,.\]
%%For details, see \cite{DyckerhoffM10a}.

Now, let a sequence $(\tilde \bma^n)_{n\in \IN}$ of independent random vectors be given that are identically distributed with $P$, and consider the sequence of random uncertainty sets ${\cal U}_n$ based on $\tilde \bma^1, \dots, \tilde \bma^n$. \cite{DyckerhoffM10} have shown:
\begin{proposition}[\cite{DyckerhoffM10}]
\label{propcont}
${\cal U}_n$ converges to  ${\cal U}$ almost surely in the Hausdorff sense.
\end{proposition}

The proposition implies that by drawing an independent sample of $\tilde \bma$ and solving the robust LP based on the observed empirical distribution a consistent estimate of the uncertainty set ${\cal U}$ is obtained. Moreover, the cutting point $\bma^{j_*}$, where the line through the origin and $\bmc$ hits the uncertainty set, is consistently estimated by our algorithm.
But, in particular for a discretely distributed $\tilde \bma$, the optimal solution $\bmx^*$ need not be a consistent estimate, as it may perform a jump when $\bma^{j_*}$ moves from one facet of ${\cal U}$ to neighboring one.

\section{Concluding remarks}
\label{secdiscuss}
A stochastic linear program (SLP) has been investigated, where the coefficients of the linear restrictions are random.
Risk constraints are imposed on the random side conditions and an equivalent robust SLP is modeled, whose worst-case solution is searched over an uncertainty set of coefficients.
If the risk is measured by a general coherent distortion risk measure, the uncertainty set of a side condition has been shown to be a \emph{weighted-mean region}. This provides a comprehensive visual and computable characterization of the uncertainty set.
An \textit{algorithm} has been developed that solves the robust SLP under a single stochastic constraint, given an external sample.
It is available as an R-package \textit{StochaTR} \citep{StochaTRpack}.
Moreover, if the data are generated by an infinite i.i.d.\ sample, the limit behavior of the solution has been investigated.
The algorithm allows the introduction of
additional deterministic  constraints, in particular, those regarding nonnegativity.

\begin{table}[h!]
    \centering
\begin{tabular}{ |c||c|c|c|c|c|c|c|c|c| }
  \hline
 $d$\textbackslash$n$ & 1000 & 2000 & 3000 & 4000 & 5000 & 10000 & 15000 & 20000 & 25000 \\ \hline
3 & 0.3 & 1.14 & 1.76 & 2.92 & 3.41 & 6.18& 12.61 & 15.06 & 47.54\\
 4 & 0.66 & 2.21 & 3.47 & 4.48 & 4.27 &7.68 & 16.97 & 20.04 & \\
 5 & 1.85 & 3.09 & 5.68 & 9.28 & 11.03 & 13.52& 27.34 & 54.86 & \\
 6 & 2.08 & 4.41 & 5.62 & 14.99 & 18.73 &25.07 & 46.88 & & \\
 7 & 2.16 & 6.22 & 13.3 & 25.44 & 28.56 & 52.33& & & \\
 8 & 4.18 & 9.78 & 20.18 & 31.82 & 34.23 & & & & \\
 9 & 5.18 & 14.75 & 24.11 & 35.94 & 61.14 & & & & \\
 10 &6.17 & 16.97 & 33.82 & 42.11 & 67.06 & & & & \\ \hline
\end{tabular}
\caption{Running times of \emph{StochaTR} for different $n$ and $d$ (in seconds).}
\label{tabcompres}
 \end{table}

Table~\ref{tabcompres} reports simulated running times (in seconds) of the R-package for the $5\%$-level expected shortfall and different $d$ and $n$. The data are simulated by mixing the uniform distribution on a $d$-dimensional parallelogram with a multivariate Gaussian distribution.
In light of the table the complexity seems to grow with $d$ and $n$ slower than ${\cal O}(d^2n^2)$. 

Besides this, we contrast our new procedure with the seminal approach of \cite{rockur00}, who solve the portfolio problem by optimizing the expected shortfall with a simplex-based method.
In illustrating their method, they simulate three-dimensional normal returns having specified expectations and covariance matrices.
We have applied our package to likewise simulated data. The results are exhibited in Table~\ref{tabcompres2}. For a comparison, some cells contain also a second value, which corresponds to the \cite{rockur00} procedure and is taken from Table 5 there. 

\begin{table}[h!]
    \centering
\begin{tabular}{ |c||c|c|c|c|c|c|c|c|c| }
  \hline
 $\alpha$\textbackslash$n$ & 1000 & 5000 & 10000 & 15000 & 20000 & 25000 \\ \hline
0.10 & 1.1 \;($<$5)& 7.2 \; (6) &23.7\; (20) &46 & 56.3\; (45)& 74.4\\
0.05 & 0.5\; ($<$5)& 4.7\; (6) &14.0\; (12) & 20.0& 39.8\; (40)& 53.2\\
0.01 & 0.3\; ($<$5) &2.3\; (6) &3.8\; (6) &7.9 &22.1\; (50) & 38.5 \\
\hline
\end{tabular}
\caption{Running times of \emph{StochaTR} for different $n$ and $\alpha$ (in seconds); in parentheses running times of \cite{rockur00}.} 
\label{tabcompres2}
 \end{table}

As we see from Table~\ref{tabcompres2}, the computational times of the two approaches do not much differ. However, our algorithm usually needs some dozens of iterations only, which is substantially less than the algorithm of \cite{rockur00}. Also, in contrast to the latter, where the resulting portfolio can vary between $(0.42,0.13,0.45)$ for $n=1000$ and $(0.64,0.04,0.32)$
% depending on the number of simulated points
for $n=5000$, we get a \textit{stable} optimal portfolio. Our solution averages at $(0.36,0.15,0.49)$, which has approximately the same $\VaR$  and expected shortfall as that in the compared study but yields a \textit{better} value of the \textit{expected return}.

Finally, our approach turns out to be very flexible. 
In particular, non-sample information can be introduced into the procedure in an interactive way
by explicitly changing and modifying the uncertainty set.
More research is needed in extending the algorithm to solve SLPs with multiple constraints \eqref{jointRiskConstraint}. Also procedures that allow for a stochastic right-hand side in the constraints and a random coefficients in the goal function have still to be explored.

\nocite{Delbaen02}
\nocite{bental00}

\section*{Acknowledgments}
Pavel Bazovkin was partly supported by a grant of the German Research Foundation (DFG).

%%\checkcomments
\bibliographystyle{ormsv080}

\bibliography{oratr}

%\bibliographystyle{plain}

%\bibliography{litmos}

\end{document}